\documentclass[11pt]{article}
\usepackage{amsmath}
\usepackage{amsthm,amssymb}

\usepackage{natbib}
\usepackage{multirow}
\usepackage{comment}

\usepackage[pdftex]{graphicx}
\usepackage{makecell}
\usepackage{enumerate}
\usepackage{booktabs}

\usepackage{array}

\usepackage{fullpage}

\usepackage{url}

\usepackage{algorithm}
\usepackage{algpseudocode}
\usepackage{bm}

\usepackage{graphicx}
\usepackage{subcaption}

\newtheorem{theorem}{Theorem}

\usepackage{mathtools}

\usepackage{wrapfig}
\usepackage{lipsum}

\usepackage{mathrsfs}
\usepackage{dsfont}
\usepackage{titling}
\usepackage{epstopdf}

\usepackage{natbib}
\usepackage{multirow}
\usepackage[usenames,dvipsnames,svgnames,table]{xcolor}
\usepackage[colorlinks=true,
            linkcolor=blue,
            urlcolor=blue,
            citecolor=blue]{hyperref}

\usepackage{amsthm}
\newtheorem{lemma}{Lemma}

\newcommand{\bbeta}{\bm{\beta}}
\newcommand{\bgamma}{\bm{\gamma}}

\newcommand{\bSigma}{\bm{\Sigma}}

\newcommand{\bC}{\bm{C}}

\newcommand{\bI}{\bm{I}}

\newcommand{\bP}{\bm{P}}

\newcommand{\bS}{\bm{S}}

\newcommand{\bU}{\bm{U}}

\newcommand{\bX}{\bm{X}}

\newcommand{\bx}{\bm{x}}

\usepackage{natbib}

\begin{document}

\title{\huge Optimal Sampling for Generalized Linear Model under Measurement Constraint with Surrogate Variables}

\author{Yixin Shen\thanks{Department of Statistics and Data Science, Cornell University, Ithaca, NY 14850, USA; e-mail: \texttt{ys964@cornell.edu}.}~~~~~Yang Ning\thanks{Department of Statistics and Data Science, Cornell University, Ithaca, NY 14850, USA; e-mail: \texttt{yn265@cornell.edu}.} 
}

\date{\today}

\maketitle

\vspace{-0.5in}

\begin{abstract}
Measurement-constrained datasets, often encountered in semi-supervised learning, arise when data labeling is costly, time-intensive, or hindered by confidentiality or ethical concerns, resulting in a scarcity of labeled data. In certain cases, surrogate variables are accessible across the entire dataset and can serve as approximations to the true response variable; however, these surrogates often contain measurement errors and thus cannot be directly used for accurate prediction. We propose an optimal sampling strategy that effectively harnesses the available information from surrogate variables. This approach provides consistent estimators under the assumption of a generalized linear model, achieving theoretically lower asymptotic variance than existing optimal sampling algorithms that do not use the surrogate data information. Using the optimal A criterion from optimal experimental design, our strategy maximizes statistical efficiency. Numerical studies demonstrate that our approach surpasses existing optimal sampling methods, exhibiting reduced empirical mean squared error and enhanced robustness in algorithmic performance. These findings highlight the practical advantages of our strategy in scenarios where measurement constraints exist and surrogates are available.

\end{abstract}
%\keyword


%\textbf{\theoremstyle{definition}
%\newtheorem{exmp}{Example}[section]}

\section{Introduction}

Semi-supervised learning is a machine learning technique that leverages both labeled and unlabeled data during training, making it especially useful when data labeling is costly, time-intensive or facing ethical issues. Such datasets, often termed "measurement-constrained datasets", are common in many fields. For example, the critical temperature of superconductors, which depends on their chemical composition, is a key property but difficult to predict due to the lack of an accurate scientific model. A data-driven approach is needed to develop materials with higher critical temperatures. Since synthesizing materials is costly and time-consuming, only a few compounds can be tested \citep{hamidieh2018}. Another example is galaxy classification. This is essential in astronomy but challenging due to the rapid growth of astronomical datasets from advanced telescopes. Since human visual classification is time-consuming and costly, it’s crucial to select a representative sample of galaxies for accurate human classification \citep{banerji2010}. 
\\~\\
Many studies on semi-supervised learning prioritize algorithm development, often overlooking statistical estimation, which poses challenges for model interpretability \citep{chapelle2010semi}. When considering statistical estimation, methodologies typically rely on nonparametric estimation techniques that can be computationally intensive for large-scale datasets \citep{cai2020semisupervised,azriel2022semi,deng2023optimal}. A widely used approach to mitigate these challenges is sampling. Rather than simply selecting data at random for labeling, this approach involves selecting a subset of data and obtaining the outcome for it based on a  sampling distribution. Sampling methods have been extensively studied, with significant developments in leverage sampling \citep{ma2015, wang2019, ma2022}, A-optimal sampling \citep{wang2018, ting2018, ai2021}, and optimal experimental design \citep{wang2017, meng2021}. %(Wang et al., 2017).
However, most of these approaches assume that response data is accessible across the full dataset, which restricts their applicability in settings where response measurements are constrained or costly to obtain. \cite{zhang2021} address this limitation by proposing a sampling method that provides statistically consistent and asymptotically normal estimators, with minimal performance compromise relative to fully labeled methods \citep{wang2018, ma2015}.
\\~\\
Another attempt to deal with the scarcity of the response variable in a measurement constrained dataset is to create a surrogate variable by various means including machine learning methods and Monte Carlo simulations. One of the important applications with this technique applied is the electronic health records (EHR). EHR are a valuable resource for health research, often used to identify novel disease risk factors. Due to a lack of the binary phenotypes of patients, researchers use phenotyping algorithms to make predictions for this phenotype as substitutions to the true conditions of patients. Even though many papers \citep{oh2021, yang2023, weinstein2023} improve the performance of the phenotyping methods, they may still misclassify patient conditions. This misclassification can introduce systematic bias, increase type I errors, and reduce statistical power in studies. \cite{tong2020} proposed a method that incorporates both true phenotype data from a validation set and phenotyping predictions across the dataset, demonstrating that this approach yields a consistent estimator with lower variance compared to methods limited to the validation-set phenotype data.
\\~\\
Building on the concepts of sampling and surrogate variable utilization as introduced by \cite{tong2020}, we propose an enhanced sampling method that preserves key statistical properties, such as statistical consistency and asymptotic normality, seen in  \cite{zhang2021}, while offering significant improvements. The primary advantage of our method lies in its integration of surrogate variable information, resulting in theoretically lower variance. This integration allows our approach to leverage contextual insights from unlabeled data, refining estimations in measurement-limited scenarios. By incorporating surrogate variables in the sampling method, our method yields more accurate estimations and enhances the performance in real-world applications. Numerical studies demonstrate that our method achieves lower empirical mean squared error, aligning well with theoretical expectations, and exhibits superior algorithmic stability compared to \cite{zhang2021}, whose estimators may diverge under certain data distributions. The consistent stability of our method across diverse datasets and sampling scenarios highlights its applicability and robustness, addressing typical scalability issues associated with semi-supervised learning and supporting its use in a range of practical contexts.

\bigskip
%%When only a small portion of a dataset can obtain responses, Y, the dataset is called a measurement-constrained dataset. This type of dataset is common in practice, for example, when the responses are expensive or time-consuming to acquire. On the other hand, even though Y is hard to gain, for many of the cases there exists a surrogate variable S, which is highly related to Y, can be attained for the whole dataset. We list several typical examples as motivations.

\section{Set up}

Let $Y$ and $\bX$ denote the response variable and a vector of $d$-dimensional covariates. In many applications, the gold-standard response variable $Y$ is not fully observable across the dataset due to cost constraints or ethical limitations. Instead, a surrogate variable, denoted by $S$, may be observed or generated, though it may be subject to measurement or classification error relative to $Y$.
\\~\\
Assume that, conditional on $\bX$, $Y$ follows a generalized linear model (GLM) with a canonical link function:
$$
f(Y|\bX;\bbeta_0)=\exp(\frac{y\bbeta_0^T\bX-b_1(\bbeta_0^T\bX)}{a_1(\phi_1)})c_1(Y),
$$
and we further impose a working model for $S$ given $\bX$, such as
$$
f(S|\bX;\bgamma_0)=\exp(\frac{S\bgamma_0^T\bX-b_2(\bgamma_0^T\bX)}{a_2(\phi_2)})c_2(S),
$$
where $a_1(\cdot)$, $a_2(\cdot)$, $b_1(\cdot)$ and $b_2(\cdot)$ are known functions; $\phi_1$ and $\phi_2$ denote known dispersion parameters; let  $\bbeta_0$, $\bgamma_0\in \mathbb{R}^p$ be the unknown parameter of interest, which is assumed to reside within compact sets $B,C \subseteq \mathbb{R}^p$. Without loss of generality, we set $ a_1(\phi_1)=a_2(\phi_2) = 1$.
\\~\\
It is worth noting that the above model for $S$ given $\bX$ may be theoretically misspecified. As in \cite{fahrmexr1990mle}, even though S is misspecified, the maximum likelihood estimators retain statistical consistency and asymptotic normality properties. The variable $S$ may follow other parametric working models or we may incorporate transformed covariates and interaction terms within the GLM framework, implying that $\bgamma_0$ is not necessarily confined to $\mathbb{R}^p$. We adopt the GLM for $S$ as mentioned above for simplicity.
\\~\\
This paper aims to develop an optimal sampling strategy for estimating $\bbeta_0$ when the true response is limited and the surrogate variable is subject to measurement or misclassification error. In particular, assume that we observe an extensive dataset with $N$ independently identically distributed (i.i.d) samples $(S_1,\bX_1),...,(S_N,\bX_N)$. Due to cost constraints, ethical considerations, or other factors, only a subset of $n$ samples (on average) can be chosen to obtain their gold-standard responses, where $n$ is often much smaller than the total sample size, $N$. Our objective is to determine an optimal sampling probability $\pi_i$ for the $i$th sample, such that we collect the subsample ${(S_i, \bX_i,Y_i)}_{i:R_i=1}$ for a Bernoulli variable $R_i$ with $P(R_i = 1|S_i, \bX_i) = \pi_i$ and $P(R_i = 0|S_i, \bX_i) = 1-\pi_i$, for $1\leq i\leq N$,where $\sum_i\pi_i = n$. The estimator of $\bbeta_0$ based on the sampled data set ${(S_i, \bX_i,Y_i)}_{i:R_i=1}$ is most statistically efficient in a class of estimators. Note that the sampling probability $\pi_i$ can only depend on the surrogate variables and covariates, as the  responses $(Y_1,...,Y_N)$ are unobserved prior to sampling.
\\~\\
A natural approach to estimating $\bbeta_0$ and $\bgamma_0$ is to derive these estimates from the solutions of their corresponding score functions. Our methodology is built upon this foundational concept, utilizing the score functions to obtain consistent and efficient estimators. A clear procedure is elaborated in the next section. 
\section{Algorithm and Asymptotic Properties}
\subsection{General Algorithm}

Here we list the general procedure of response-free sampling with surrogates in Algorithm 1. Solving a weighted score equation in step 2 guarantees the unbiasedness of the estimating equation \citep{zhang2021}. The formulation of the augmented estimator in step 4 is inspired by \cite{tong2020}, which can be obtained by projecting $\hat\bbeta_n-\bbeta$ onto  $\hat\bgamma_n-\hat\bgamma_N$. Due to the orthogonality of $\hat{\bbeta}_A-\bbeta$ and $\hat\bgamma_n-\hat\bgamma_N$, the augmented estimator $\hat{\bbeta}_A$ is asymptotically more efficient than 
$\hat\bbeta_n$.

\begin{algorithm}[H]
\caption{Response-free Optimal Sampling for GLM with Surrogates}

\begin{algorithmic}[1]

\State Generate $R_i$ with $P(R_i = 1|S_i, \bX_i) = \pi_i$ and $P(R_i = 0|S_i, \bX_i) = 1-\pi_i$, for $1\leq i\leq N$. Get the subsample ${(S_i, \bX_i,Y_i)}_{i:R_i=1}$. %{\color{red}Sample with replacement} from N observations n times with weights $\{\pi_i\}^N_{i = 1}$. $\pi_i$ only depends on $(X_1,\dots,X_N)$, $(S_1,\dots,S_N)$ and a pilot estimate of $\beta$, but not $(Y_1,\dots,Y_N)$. Get the subsample ${(X_i^*,Y_i^*)}_{i=1}^n$.

\State Given the sampling probability $\pi_i$, we estimate $\bbeta_0$ and $\bgamma_0$ by solving the re-weighted score equations 
$$
\sum_{i=1}^N\frac{R_i}{\pi_i}(Y_i-b_1'(\bbeta_0^T\bX_i))\bX_i=0,
$$

$$
\sum_{i=1}^N\frac{R_i}{\pi_i}(S_i-b_2'(\bgamma_0^T\bX_i))\bX_i=0,
$$
where $R_i=1$ if the $i$th sample is selected, and $R_i=0$ otherwise.  The estimators are denoted by $\hat\bbeta_n$ and $\hat\bgamma_n$.

\State Solve the score equation for $\bgamma_0$ based on the entire data set

$$
\sum_{i=1}^N(S_i-b_2'(\bgamma_0^T\bX_i))\bX_i=0.
$$
The estimator is denoted by $\hat\bgamma_N$. 
\algstore{myalg}
\end{algorithmic}
\end{algorithm}
\begin{algorithm} [H]                   
\begin{algorithmic}[1]               
\algrestore{myalg}
\State Construct the final augmented estimator 
$$
\hat\bbeta_A=\hat\bbeta_n-\hat\bSigma_{12}\hat\bSigma_{22}^{-1}(\hat\bgamma_n-\hat\bgamma_N),
$$ where we define $\bSigma$$ 
= \begin{pmatrix} 
\bSigma_{11} & \bSigma_{12} \\
\bSigma_{21} & \bSigma_{22}
\end{pmatrix}$
 as the asymptotic covariance matrix of $n^{1/2}(\hat\bbeta_n-\bbeta, \hat\bgamma_n-\hat\bgamma_N)$, and $\hat\bSigma_{12},\hat\bSigma_{22}$ are the estimates of the corresponding block matrices. 
\end{algorithmic}
\end{algorithm} 

In practice, after selecting a subsample of size $n$, only these $n$ responses are required to be measured, which can lead to substantial cost reductions. This is especially beneficial in scenarios where response measurements are resource-intensive or costly, as it allows for efficient allocation of limited resources. The success of this approach, however, is closely linked to the determination of sampling weights $\pi_i$ and the appropriate subsample size $n$. Under constraints imposed by measurement resources, the subsample size $n$ is generally determined by the cost of collecting response measurements, as well as the available computational and storage capacity. The choice of sampling weights $\pi_i$ is critical for achieving an efficient estimation process. We employ a data-driven approach to derive a sampling distribution for $\pi_i$ that maximizes the estimator's efficiency, which will be elaborated in the subsequent sections.
%\\~\\
%We demonstrate that $\hat{\bbeta}_A$ possesses several important statistical properties:

\subsection{Consistency of $\hat{\bbeta}_A$}
The following theorem shows the consistency of $\hat{\bbeta}_A$.
\setcounter{theorem}{0}
\begin{theorem}
If:
  \begin{enumerate}[(i)]
    \item $(ia)$ $b_1''(\cdot)$, $b_2^{''}(\cdot)$, or $(ib)$$\bX$ is almost surely bounded.   \label{Con1}
    
    \item $E\bX\bX^T$ is finite,  $g_1(\bbeta):=E\left[\{b_1'(\bX^T\bbeta)-Y\}\bX\right]$ is finite for any $\bbeta\in \mathcal{B}$, and $g_2(\bgamma):=E\left[\{b_2'(X^T\bgamma)-S\}X\right]$ is finite for any $\bgamma\in \mathcal{A}$, where $\mathcal{A}$ and $\mathcal{B}$ are compact sets. \label{Con2}
    
    \item $\sum_{i=1}^N E\left[ \frac{\{b_1'(\bX_i^T\bbeta)-Y_i\}^2}{\pi_i} x_{ij}^2\right]=o(N^2n)$ , where $1 \leq j \leq p$,\\ and $\bbeta \in \mathcal{B}$,
    and $\sum_{i=1}^N E\left[ \frac{\{b_2'(\bX_i^T\bgamma)-Y_i\}^2}{\pi_i} x_{ij}^2\right]=o(N^2n)$ for $1\le j\le p$ and $\bgamma \in \mathcal{A}$. \label{Con3}
    
    \item $\inf_{\bbeta:||\bbeta-\bbeta_0||\ge \epsilon_1}||E[(b_1'(\bX^T\bbeta) - Y)\bX]||>0$ for any $\epsilon_1>0$, and $\inf_{\bgamma:||\bgamma-\bgamma_0||\ge \epsilon_2}||E[(b_2'(\bX^T\bgamma) - S)\bX]||>0$ for any $\epsilon_2>0$.\label{Con4}
        
    \item $\bSigma_{22} = Var(\hat\bgamma_n-\hat\bgamma_N)$  strictly positive definite, and $\bSigma_{22} $ and $\bSigma_{12} =  Cov(\hat\bbeta_n, \hat\bgamma_n-\hat\bgamma_N)$ are finite.\label{inver}

    \item $E[|\bX_i||^4]$ is finite for $1<i<N$. \label{consis}

 \end{enumerate}

\noindent then we have $\hat{\bbeta}_A \mathop{\to}\limits^{p}\bbeta$.

\end{theorem}

Similarly as in \cite{zhang2021}, condition (\ref{Con1}$a$) is satisfied for most GLM except Poisson regression. For Poisson regression, the theoretical framework is applicable if Condition (\ref{Con1}b) is fulfilled. To apply a consistency theorem for M-estimators \citep{vandervaart2000}, Conditions (\ref{Con3}) and (\ref{Con4}) are necessary. Condition (\ref{Con3}) guarantees the uniform convergence of $S_n^*$, while Condition (\ref{Con4}) is a standard \emph{well-separated} condition for consistency proofs. This condition is met if $g_1(\cdot)$ and $g_2(\cdot)$ have unique minimizers.
\bigskip

\subsection{Asymptotic Normality $\hat{\bbeta}_A$}
For clarification, we denote the score function
$$
S_N(\bC)=
\sum_{i=1}^N(S_i-b'(\bC^T\bX_i))\bX_i,
$$
and re-weighted score function 
$$
S^*_N(\bC)=
\sum_{i=1}^N\frac{R_i}{\pi_i}(Y_i-b'(\bC^T\bX_i))\bX_i.
$$

We elaborate on the asymptotic normality of $\hat{\bbeta}_A$ through the following theorem: 
\begin{theorem}

If:
	\begin{enumerate}[(i)]
		\item The matrices $I(\bbeta_0)=-E\left\{b_1^{''}(X^T\bbeta_0)\bX\bX^T\right\}$ and $I(\bgamma_0)=-E\left\{b_2^{''}(\bX^T\bgamma_0)\bX\bX^T\right\}$ are finite and non-singular. \label{finite}
		\item $\sum_{i=1}^{N}E\left\{\frac{b_1^{''}(\bX_i^T\bbeta_0)^2}{\pi_i}(x_{ik}x_{ij})^2\right\}=o(N^2n)$ and $\sum_{i=1}^{N}E\left\{\frac{b_2^{''}(\bX_i^T\bgamma_0)^2}{\pi_i}(x_{ik}x_{ij})^2\right\}=o(N^2n)$,
        \\ and $\sum_{i=1}^{N}E\left\{\frac{b_1^{''}(\bX_i^T\bbeta_0)b_2^{''}(\bX_i^T\bgamma_0)}{\pi_i}(x_{ik}x_{ij})^2\right\}=o(N^2n)$,  for $1\le k,j \le p$. \label{finite2}
		\item $b_1(x)$ and $b_2(x)$ is three-times continuously differentiable for every $x$ within its domain.
		\item Every second-order partial derivative of $\xi_{\bbeta}(\bX) =(b_1'(\bX^T\bbeta)-Y)\cdot \bX$ w.r.t $\bbeta$ is dominated by an integrable function $\ddot{\xi}(\bX)$ independent of $\bbeta$ in a neighborhood of $\bbeta_0$, and every second-order partial derivative of $\xi_{\bgamma}(\bX)$ w.r.t $\bgamma$ is dominated by an integrable function $\ddot{\xi}(\bX)$ independent of $\bgamma$ in a neighborhood of $\bgamma_0$.  \label{AL3}
            \item $E[||\bX_i||^4]$ is finite for $1<i<N$. \label{consis2} 
	\end{enumerate}
\bigskip
We have $$\hat{\bbeta}_A - \bbeta_0 \xrightarrow{d} N(0, \bSigma_{11} - \bSigma_{12}\bSigma_{22}^{-1}\bSigma_{21}),$$
where $\bSigma_{11} = I(\bbeta_0)^{-1}\text{cov}(S^*_N(\bbeta_0),S^*_N(\bbeta_0))I(\bbeta_0)^{-1}$, $\bSigma_{12}=I(\bbeta_0)^{-1}\text{cov}(S_N^*(\bbeta_0), (S_N^*(\bgamma_0) - S_N(\bgamma_0)))$
$I(\bgamma_0)^{-1}$, $\bSigma_{22} = I(\bgamma_0)^{-1}\text{cov}((S_N^*(\bgamma_0) - S_N(\bgamma_0)), (S_N^*(\bgamma_0) - S_N(\bgamma_0)))I(\bgamma_0)^{-1}$, and $\bSigma_{21}=\bSigma_{12}^T.$
%Explicitly,
%$$\hat{\bbeta} - \bbeta \xrightarrow{d} N(0, I(\bbeta_0)^{-1}[\text{cov}(S^*_N(\bbeta_0),S^*_N(\bbeta_0))-\text{cov}(S_N^*(\bbeta_0), (S_N^*(\bgamma_0) - S_N(\bgamma_0))) $$$$\cdot\text{cov}((S_N^*(\bgamma_0) - S_N(\bgamma_0)), (S_N^*(\bgamma_0) - S_N(\bgamma_0)))^{-1}\text{cov}(S_N^*(\bbeta_0), (S_N^*(\bgamma_0) - S_N(\bgamma_0)))^T]I^{-1}(\bbeta_0)),$$

\end{theorem}

Following \cite{zhang2021}, we utilize the A-optimality criterion for experimental design, as described by \cite{kiefer1959}. This criterion involves minimizing the trace of the matrix expression \(\text{trace}(\bSigma_{11} - \bSigma_{12}\bSigma_{22}^{-1}\bSigma_{21})\), which is directly related to the minimization of the asymptotic mean squared error. In practice, this means identifying an optimal sampling distribution \(\pi_i\) that minimizes this trace.

\subsection{Weights under Constraints with Surrogates}

In the last expression above, each $S_N^*(\cdot)$ contains $\pi$, and with the inverse of $\text{cov}((S_N^*(\bgamma_0) - S_N(\bgamma_0)), 
$
$(S_N^*(\bgamma_0) - S_N(\bgamma_0)))$, a closed form of $\pi_i$ cannot be found. Instead, we find the $\pi_i$ that can minimize \\
$ I(\bbeta_0)^{-1}\text{cov}(S^*_N(\bbeta_0),S^*_N(\bbeta_0))I^{-1}(\bbeta_0)$. Notice that compared to \cite{zhang2021}, the variance is conditioned on a larger vector space $\{\bX, \bS\}$ than $\{\bX\}$. So the minimized value of \\$ I(\bbeta_0)^{-1}\text{cov}(S^*_N(\bbeta_0),S^*_N(\bbeta_0))I^{-1}(\bbeta_0)$ would be smaller than the variance shown in \cite{zhang2021}. We also show in the Appendix that 
$$I(\bbeta_0)^{-1}\text{cov}(S^*_N(\bbeta_0),S^*_N(\bbeta_0))I^{-1}(\bbeta_0)\geq I(\bbeta_0)^{-1}[\text{cov}(S^*_N(\bbeta_0),S^*_N(\bbeta_0))-\text{cov}(S_N^*(\bbeta_0), (S_N^*(\bgamma_0) - S_N(\bgamma_0))) $$$$\cdot\text{cov}((S_N^*(\bgamma_0) - S_N(\bgamma_0)), (S_N^*(\bgamma_0) - S_N(\bgamma_0)))^{-1}\text{cov}(S_N^*(\bbeta_0), (S_N^*(\bgamma_0) - S_N(\bgamma_0)))^T]I^{-1}(\bbeta_0),$$ so the overall asymptotic variance would even smaller. 
\\~\\
We can now obtain that 

\begin{theorem}
    
When $$\pi_i \propto \sqrt{E(Y^2| S_i, \bX_i) - 2E(Y|S_i, \bX_i)b_1'(\bbeta_0^T\bX_i)+b_1'^2(\bbeta_0^T\bX_i)}||I(\bbeta_0)^{-1}\bX_i||_2,$$

trace$(\bSigma_{11})$ = trace$(I(\bbeta_0)^{-1}\text{cov}(S^*_N(\bbeta_0),S^*_N(\bbeta_0))I^{-1}(\bbeta_0)\geq I(\bbeta_0)^{-1})$  can be minimized. 
\end{theorem}

Notice that here we need to use some non-parametric method like random forest to make predictions for $E(Y^2| S_i, \bX_i)$ and $E(Y|S_i, \bX_i)$. So this method would give a more accurate result in application when the test MSE is smaller. Here people can use any non-parametric method, and there may exist some methods that give better results than the random forest. Also, the optimal weights cannot be computed directly in practice because they rely on the population-level quantities \(I^{-1}\) and \(\bbeta_0\). As a result, to carry out response-free sampling, we need preliminary estimates of \(I\) and \(\bbeta_0\).
\\~\\
The following is a more detailed algorithm:

\begin{algorithm}[H]
\caption{Optimal Sampling under Measurement Constraint with Surrogates (OSUMCS)}\label{getgpx}
\begin{algorithmic}

\State 
1. First randomly sample $n_0$ data ($n_0$ $\ll$ n) and acquire their responses so that we can fit GLMs to obtain initial estimators $\hat\bbeta_n$, $\hat\bgamma_n$ and $\hat\bgamma_N$. 
\bigskip

\State
2. Calculate the estimator $\hat{I} = -\frac{1}{n_0}\sum^{n_0}_{i = 1} b_1''(\hat{\bbeta}_0^T\bX_i) \bX_i\bX_i^T$, and by using random forest (or some other non-parametric method) to get estimates for $E(Y^2_i| S_i, \bX_i)$ and $E(Y_i|S_i, \bX_i)$.
\bigskip

\State 
3. Get $$\pi_i \propto \sqrt{E(Y^2| S_i, \bX_i) - 2E(Y|S_i, \bX_i)b_1'(\bbeta_0^T\bX_i)+b_1'^2(\bbeta_0^T\bX_i)}||I(\bbeta_0)\bX_i||_2,$$ for all i. 

\bigskip

\State 
4. Apply Algorithm 1 with $\pi_i$ to get $\hat{\bbeta}_A.$

\end{algorithmic}
\end{algorithm}
Notice that step 1 in the above algorithm is designed to provide initial estimations in situations where few response values are available initially, or collecting some responses is feasible despite associated costs. In cases where a moderate number of responses are already accessible in the initial data pool, these existing values can be utilized to compute pilot estimators. If no initial responses are available, an alternative approach is to draw a small random sample from the data, using uniform sampling probabilities to create a pilot subset. 
\\~\\
The size of the pilot sample, denoted as $n_0$, should be relatively small compared to the total sample size N. In our empirical study, we chose $n_0 = 500$ for a dataset with $N = 10^5$ and dimensionalities p varying between 20 and 100, demonstrating that this algorithm performs effectively under these conditions.
\\~\\
In practice, we need $\sqrt{E(Y^2| S_i, \bX_i) - 2E(Y|S_i, \bX_i)b_1'(\bbeta_0^T\bX_i)+b_1'^2(\bbeta_0^T\bX_i)}$ to be non-negative, so we use the random forest to estimate the whole $\sqrt{E(Y^2| S_i, \bX_i) - 2E(Y|S_i, \bX_i)b_1'(\bbeta_0^T\bX_i)+b_1'^2(\bbeta_0^T\bX_i)}$ in numeric studies and the results are non-negative in all of our cases. In cases where this estimate becomes negative, one can establish a threshold slightly above zero to replace any negative values, ensuring all estimates remain non-negative. 

\section{Simulations}

Here we show our methodology and compare the results on simulated data. All works are run in a Python environment on a Mac laptop with 8 GB RAM with processor 2.4 GHz Quad-Core Intel Core i5.

\subsection{Logistic Regression}
%alter_pi_log_linear_wrf.py
%paper_result_pi/tao/avg_1/2_log_linear_wrf.txt
%paper_result_pi/tao/avg_3/4/5/6_log_linear_wrf_c.txt
Recall for logistic regression, we have $\bP(Y = 1 | \bX, \bbeta_0) = 1- \frac{1}{1 + e^{\bbeta_0^T\bX}}$, and $b_1(\bbeta_0^T\bX) =log(1+e^{\bbeta_0^T\bX}) $, thus $b_1'(\bbeta_0^T\bx) = 1-\frac{1}{1+e^{\bbeta_0^T\bx}}$ and $b_2''(\bbeta_0^T\bX) =\frac{e^{\bbeta_0^T\bX}}{(1+e^{\bbeta_0^T\bX})^2}$. Here we sample N = 100,000 under logistic models. Let $\bbeta_0 = (\underbrace{0.5,...,0.5}_\text{10}) $ and $S = Y\zeta + 5\bX\bbeta_0 + \bX\eta + \epsilon$, where $\zeta \sim N(5, 0.04\bI)$, $\eta \sim N(0, 0.25\bI)$, and $\epsilon \sim N(0, 0.25\bI)$.
\\
Our method tends to work better when there are fewer outliers in the outcome, because if it is excessive, the predictions for $E(Y^2| S_i, \bX_i)$ and $E(Y|S_i, \bX_i)$ are hard to have relatively small test MSE, then it is unlikely to induce desired final results. Here we roughly follow the settings in \cite{wang2018} and \cite{zhang2021}, and consider the six scenarios for the covariate matrix X:
\bigskip
\\
1. \textbf{mzNormal.} X is generated from $N(0, \bSigma),$ where $\bSigma_{ij} = 0.5^{\mathbf{1}(i \neq j)}$, and this is a balanced dataset, which means the 0 and 1s in the response Y are almost equal. 

2. \textbf{nzNormal.} X is generated from $N(0.5, \bSigma)$, and here about 75\% of the response Y are 1s.

3. \textbf{unNormal.} X is generated from $N(0, \bSigma_1),$ where $\bSigma_1 = \bU_1\bSigma\bU_1$ and $\bU_1 = diag(1, 1/2, ...,1/10).$
\\
4. \textbf{mixNormal.} X is generated from $0.5N(0.5, \bSigma)+0.5N(-0.5, \bSigma).$
\\
5. \textbf{T3.} X is generated from a t distribution (with a degree of freedom 3), $t_3(0, \bSigma)/10.$ This distribution has relatively heavy tails, and the 0s and 1s in the response Y are almost equal. Notice here the moment assumptions in our theorems are violated, we want to see how our method performs when these assumptions are violated. 

6. \textbf{Exp.} Each covariate in X is independent and follows exp(2). The distribution is right-skewed, and 84\% of Y are 1s.

\begin{figure}[H]
    \centering
    \begin{subfigure}{0.32\textwidth}
        \includegraphics[width=\linewidth]{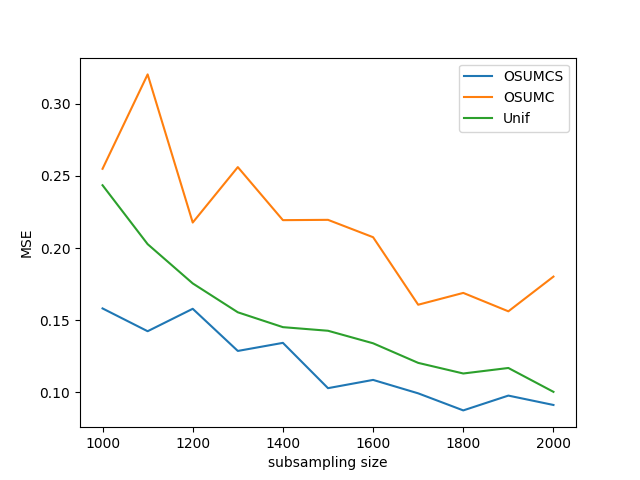}
        \caption{mzNormal}
    \end{subfigure}
    \begin{subfigure}{0.32\textwidth}
        \includegraphics[width=\linewidth]{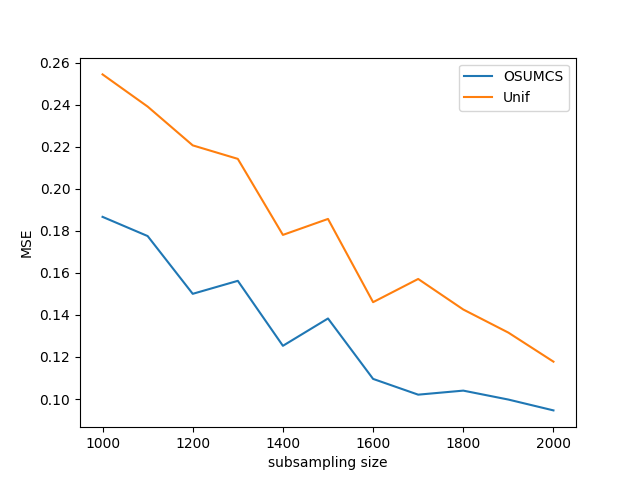}
        \caption{nzNormal}
    \end{subfigure}
    \begin{subfigure}{0.32\textwidth}
        \includegraphics[width=\linewidth]{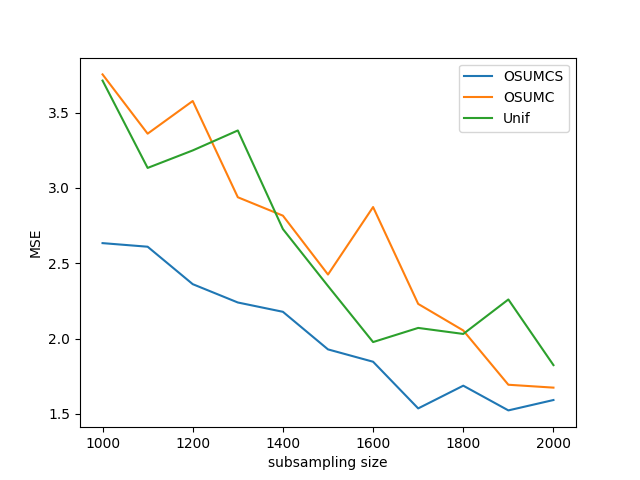}
        \caption{unNormal}
    \end{subfigure}
\end{figure}
\begin{figure}[H]\ContinuedFloat
    \centering
    \begin{subfigure}{0.32\textwidth}
        \includegraphics[width=\linewidth]{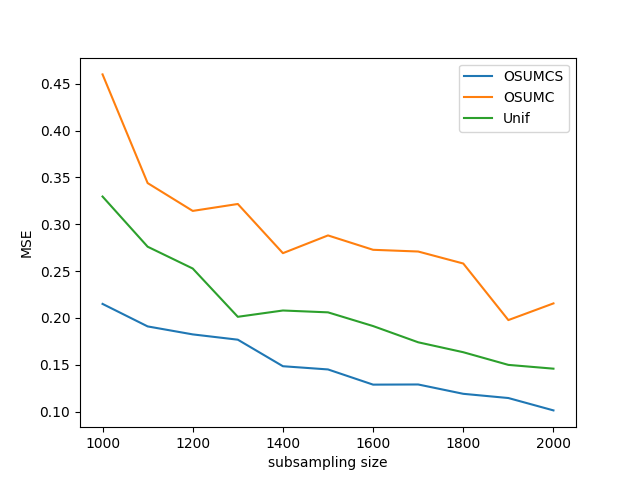}
        \caption{mixNormal}
    \end{subfigure}
    \begin{subfigure}{0.32\textwidth}
        \includegraphics[width=\linewidth]{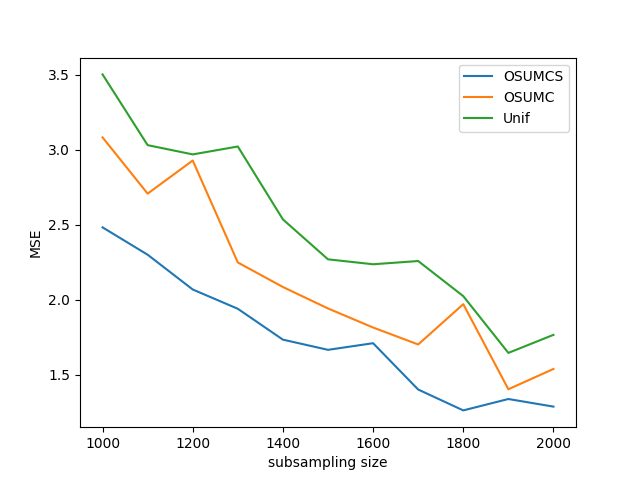}
        \caption{T3}
    \end{subfigure}
    \begin{subfigure}{0.32\textwidth}
        \includegraphics[width=\linewidth]{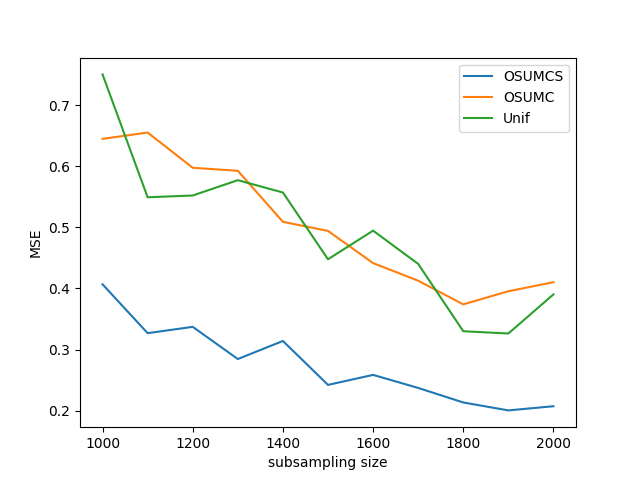}
        \caption{Exp}
    \end{subfigure}

    \caption{MSE is evaluated for the proposed optimal sampling strategy (OSUMCS), the method introduced by \cite{zhang2021} (OSUMC), and uniform sampling (Unif) across varying subsample sizes, n, under six distinct logistic regression scenarios.}
    \label{fig:six_images}
\end{figure}

For each scenario, we compare our OSUMCS result with the method by \cite{zhang2021} (OSUMC) and uniform sampling (Unif), across a range of subsample sizes, n, varying from 1000 to 2000 in increments of 100. Both OSUMCS and OSUMC employ an initial uniform sampling step with a subsample size of 
$n_0=500$ for generating pilot values. To ensure consistency in estimation, we use standard Newton's method across all three methods, setting the same pilot estimator value as the starting point for the iterative process. 
\\~\\
To assess the accuracy and consistency of each method, we run $S = 50$ simulation iterations, calculating the empirical Mean Squared Error (MSE) as $S^{-1} \sum_{i=1}^S ||\hat{\bbeta}_i - \bbeta_0||^2$, where $\hat{\bbeta}_i$ is the estimate from the $i$th iteration, and $\bbeta_0$ represents the true parameter values. The empirical MSE values obtained from these simulations are summarized in Figure 1, providing a visual comparison of the estimation accuracy across methods and subsample sizes.
\\~\\
Figure 1 demonstrates that our method, OSUMCS, consistently achieves a smaller empirical Mean Squared Error (MSE) compared to both OSUMC and Unif. This lower MSE indicates that OSUMCS produces estimates with smaller deviations from the true values and fewer large errors, highlighting its greater accuracy. This result aligns with theoretical expectations, confirming that OSUMCS has a lower variance than OSUMC.
\\~\\
In the second scenario, OSUMC encounters convergence issues during Newton's method, causing it to diverge. As a result, we present results only for OSUMCS and Unif in this case. Similar divergence issues arise in the first and third scenarios, so we report results from the 50 simulations in which OSUMC converges across all cases. This divergence suggests that OSUMCS offers improved accuracy and demonstrates enhanced robustness in terms of algorithmic stability and performance. 
\\~\\
In the fifth scenario, since the $T_k$ distribution only has moments up to order
$k-1$, this violates the moment assumptions underlying both OSUMCS and OSUMC, and Unif outperforms OSUMC. However, our OSUMCS method still surpasses Unif, demonstrating the advantages of leveraging information from the surrogate variable. This outcome underscores OSUMCS’s ability to achieve superior results under challenging conditions, reflecting the value of incorporating auxiliary information for improved estimation. 

\subsection{Linear Regression}
%alter_pi_with_rf.py
%paper_result_pi/tao/avg_#_linear_linear_wrf.txt
Let N = 100,000, $Y = \bX\bbeta_0 + \epsilon_1$, and $S =10Y + \bX\eta + \epsilon_2$, where $\epsilon_1 \sim N(0, 9\bI)$, $\eta \sim N(2,\bI)$, and $\epsilon_1 \sim N(0, \bI)$. For linear regression, we have $b_1(\bbeta_0^T\bX) =\frac{1}{2}(\bbeta_0^T\bX)^2$, thus $b_1'(\bbeta_0^T\bX) = \bbeta_0^T\bX$ and $b_1''(\bbeta_0^T\bX) = 1$. Let $\bbeta_0 = (\underbrace{0.5,...,0.5}_\text{30})$, and following \cite{wang2017}, \cite{ma2015} and \cite{zhang2021}, there are three scenarios for the covariate matrix $\bX$:
\\
1. \textbf{GA.}
X is generated from $N(\mathbf{1},\bSigma)$, where $\bSigma_{ij} = 2 \times 0.5^{|i-j|}.$\\
2. \textbf{T3.} 
X is generated from $t_3(0,\bSigma)$, where $t_3$ refers to the t-distribution with 3 degrees of freedom. Again, the moment assumptions in our theorems are violated. \\
3. \textbf{T1.}
X is generated from $t_1(0,\bSigma)$, where $t_1$ refers to the t-distribution with 1 degree of freedom. The moment assumptions in our theorems are also violated.

\begin{figure}[H]
    \centering
    \begin{subfigure}{0.32\textwidth}
        \includegraphics[width=\linewidth]{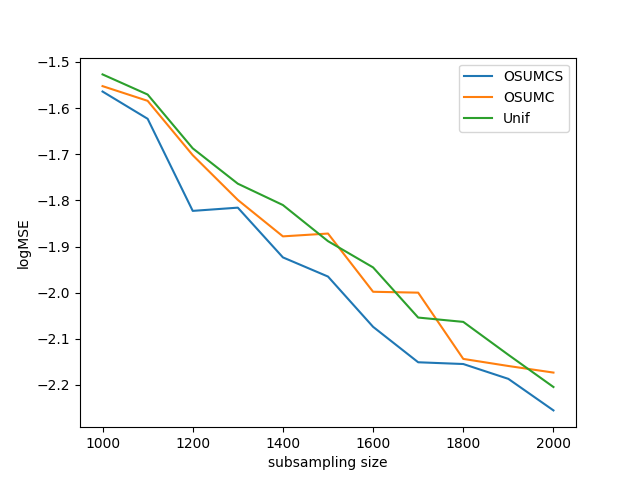}
        \caption{GA}
    \end{subfigure}
    \begin{subfigure}{0.32\textwidth}
        \includegraphics[width=\linewidth]{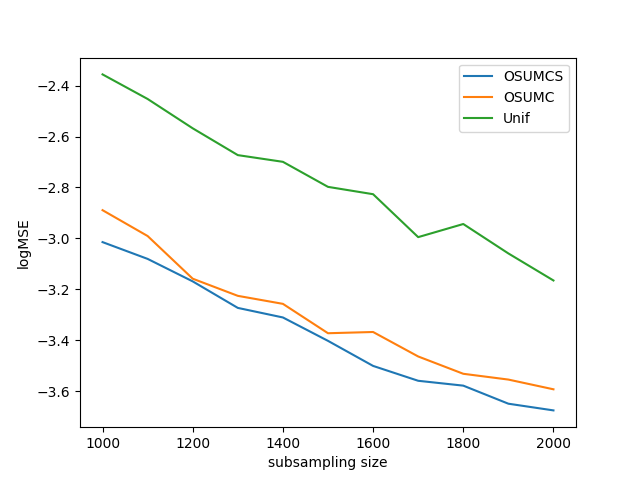}
        \caption{T3}
    \end{subfigure}
    \begin{subfigure}{0.32\textwidth}
        \includegraphics[width=\linewidth]{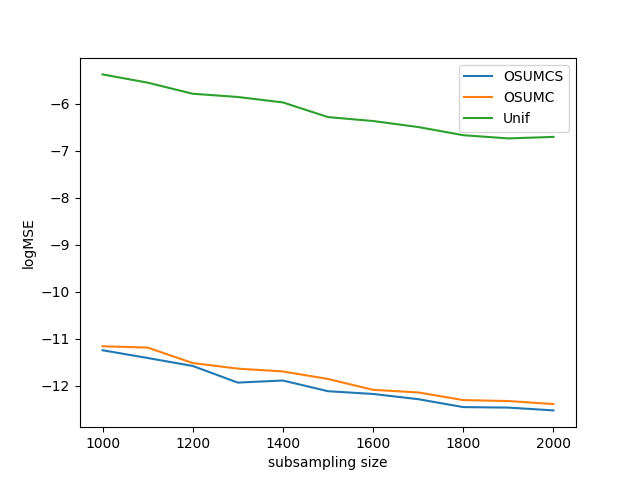}
        \caption{T1}
    \end{subfigure}

    \caption{Comparisons of logMSE results of OSUMCS, OSUMC, and Unif across varying subsample sizes, n, under three linear regression scenarios.}
\end{figure}

We assess the performance of our OSUMCS method in comparison with OSUMC and Unif across the same range of subsample sizes as those evaluated under the logistic model assumptions. Each simulation is repeated 100 times, and for clarity, we present the empirical logarithm of the MSE in Figure 2.
\\~\\
In all three design scenarios, OSUMCS consistently outperforms the other methods, resulting in lower MSE values that align closely with our theoretical predictions. In the T3 and T1 settings, even though the moment assumptions required by the theorems for OSUMCS and OSUMC are not fully met, both methods still significantly outperform uniform sampling, demonstrating robustness to certain assumption violations. In the linear design setting, the OSUMC already performs well, and the additional information from the surrogate variable in OSUMCS further enhances the accuracy of the estimation. 

\subsection{Poisson Regression}
%alter_pi_poi_linear_wrf.py
%paper_result_pi/tao/avg_#_poi_linear_wrf.txt
Let N = 100,000, $\lambda = e^{\bX\bbeta_0}$, and $S =5Y + \zeta\bX\bbeta_0 + \epsilon_1$, where $\zeta \sim N(5, 0.09\bI)$, and $\epsilon_1 \sim N(0, \bI)$. 
Under this setting, we have $b_1(\bbeta_0^T\bX) =e^{\bX\bbeta_0}$, thus $b_1'(\bbeta_0^T\bX) = e^{\bX\bbeta_0}$ and $b_1''(\bbeta_0^T\bX) = e^{\bX\bbeta_0}$. 
Let $\bbeta = (\underbrace{0.1,...,0.1}_\text{10})$, which lets $\lambda$ to have a moderate size, and following \cite{ma2015} and \cite{zhang2021}, there are four scenarios for the covariate matrix $\bX$:
\\
1. \textbf{mzNormal.} 
X is generated from $N(0, \bSigma),$ where $\bSigma_{ij} = 0.5^{\mathbf{1}(i \neq j)}$, and this is a balanced dataset, which means the 0 and 1s in the response Y are almost equal. 
\\
2. \textbf{nzNormal.}
X is generated from $N(0.5, \bSigma)$, and here about 75\% of the response Y are 1s.
\\
3. \textbf{Uniform.}
X is generated from an independent uniform distribution over $[-1, 1]$ for the first half of X, and over $[-0.5, 0.5]$ for the rest half of X.

4. \textbf{T3.}
X is generated from a t distribution (with a degree of freedom 3), $t_3(0, \bSigma)/10.$ Here the 0s and 1s in the response Y are almost equal, and the moment assumptions in our theorems are violated. 

\begin{figure}[H]
    \centering
    \begin{subfigure}{0.4\textwidth}
        \includegraphics[width=\linewidth]{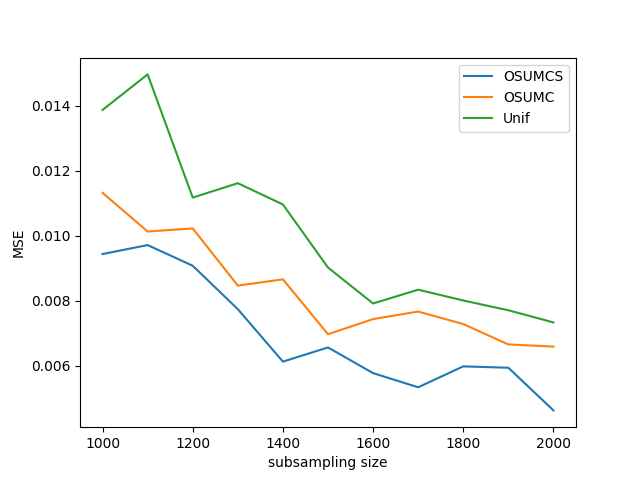}
        \caption{mzNormal}
    \end{subfigure}
    \begin{subfigure}{0.4\textwidth}
        \includegraphics[width=\linewidth]{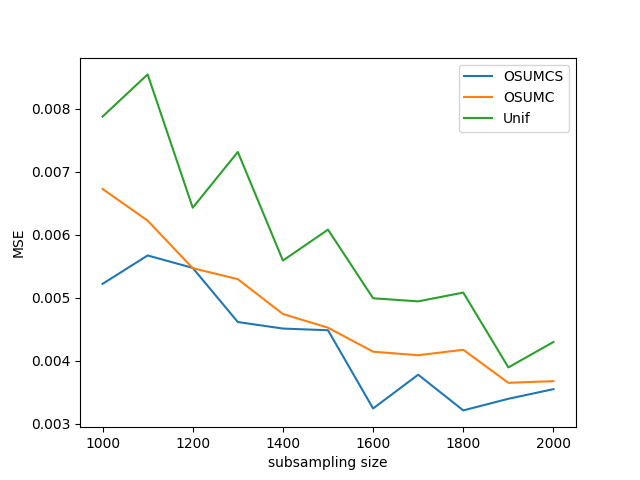}
        \caption{nzNormal}
    \end{subfigure}
\end{figure}
\begin{figure}[H]\ContinuedFloat
    \centering
    \begin{subfigure}{0.4\textwidth}
        \includegraphics[width=\linewidth]{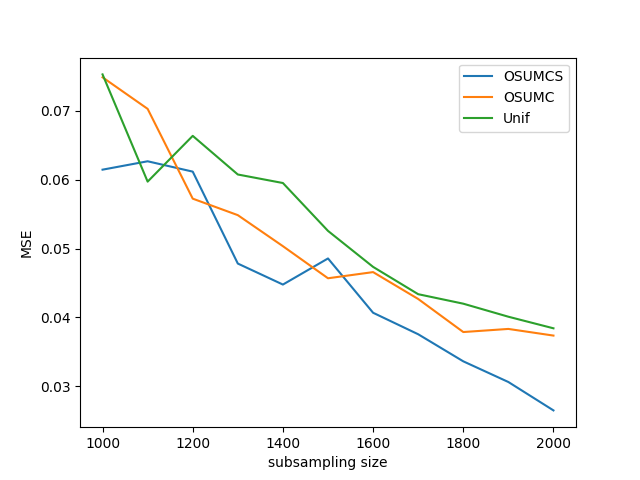}
        \caption{Uniform}
    \end{subfigure}
    \begin{subfigure}{0.4\textwidth}
        \includegraphics[width=\linewidth]{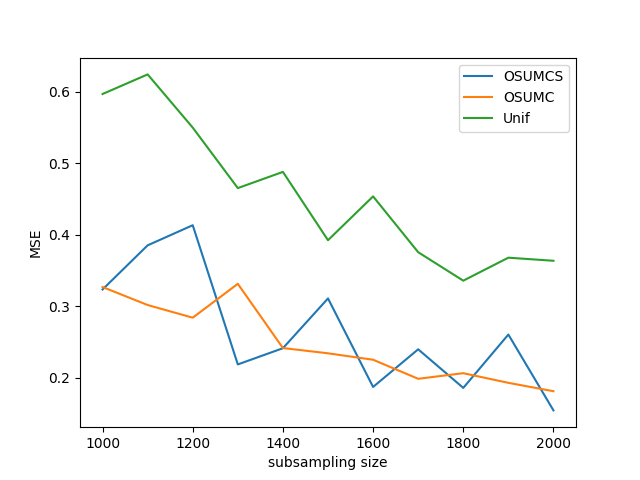}
        \caption{T3}
    \end{subfigure}

    \caption{Comparisons of MSE results of OSUMCS, OSUMC, and Unif across varying subsample sizes, n, under four poisson regression scenarios.}
\end{figure}
We evaluate the performance of our OSUMCS method alongside OSUMC and Unif over the same subsample sizes used in the previous analyses. Each simulation is repeated 50 times, and we display the empirical MSE in Figure 3.
\\~\\
In both the first and second scenarios, where the covariate matrix $X$ is generated from a normal distribution, our OSUMCS method consistently outperforms both OSUMC and Unif. However, when $X$ follows a uniform distribution, the Poisson-distributed response variable $Y$ occasionally takes on extremely large values, which introduces potential outliers. These outliers reduce the accuracy of the conditional expectations $E(Y^2| S_i, \bX_i)$ and $E(Y|S_i, \bX_i)$, bringing OSUMCS’s performance closer to that of OSUMC and Unif.
In the fourth scenario, in addition to the presence of outliers, performance is further impacted by the violation of moment assumptions. This dual effect—outliers combined with unmet assumptions—diminishes the advantage provided by the surrogate variable, leading to a similar performance to OSUMC.

\section{Application}
%alter_pi_with_rf_one_s_real_data.py
%paper_result_pi/tao/avg_linear_linear_wrf_real_data_2_woY/wY.txt
We here use the same dataset as \cite{zhang2021} used, the superconductivity data from \cite{hamidieh2018}, which is available through the UCI Machine Learning Repository at https://archive.ics.uci
.edu/ml/datasets/Superconductivty+Data. The objective of this study is to build a predictive model for the critical temperature at which materials transition to a superconducting state, based on chemical composition data. This dataset contains critical temperature values for 21,263 superconducting materials and 81 features derived from their chemical formulas. In Hamidieh’s original analysis, a multiple linear regression model was applied to the full dataset to calculate regression coefficients, which \cite{zhang2021} use as the "true" parameter $\bbeta_0$  in the evaluations, and we here also adopt this idea for comparison. 
\\~\\
To perform the comparison, we randomly split the data, using 19,000 observations as a training set and reserving the remaining data as a test set. Each sampling method is then applied to the training set to generate the coefficient estimate $\hat{\bbeta}$. We assess the performance of our OSUMCS method against OSUMC and Unif within a linear regression context. In addition to estimation accuracy, we compare the prediction capability of each sampling approach. 
\\~\\
Even though there doesn't exist a surrogate S in the dataset, we still can build S as the following. First let $$\bgamma_0 = (\bX^T\bX)^{-1}\bX^TY_{n_0},$$ where $Y_{n_0}$ comes from the small pilot sample from the initial $n_0$ generation. Define S as: $$S = 3\bX^T\bgamma_0 + \epsilon,$$ where $\epsilon \sim N(0,1).$
\\

We introduce noise to ensure that $S$ is not a direct linear combination of $\bX$, and the term $\bX^T \bgamma_0$ is scaled by a factor of 3 to maintain the relative magnitude of the noise at a manageable level. This scaling factor is adjustable, as other values could be used to achieve similar effects. 

\begin{figure}[H]
	\centering
	\begin{subfigure}[b]{0.49\linewidth}
            \centering
		\includegraphics[width=\textwidth]{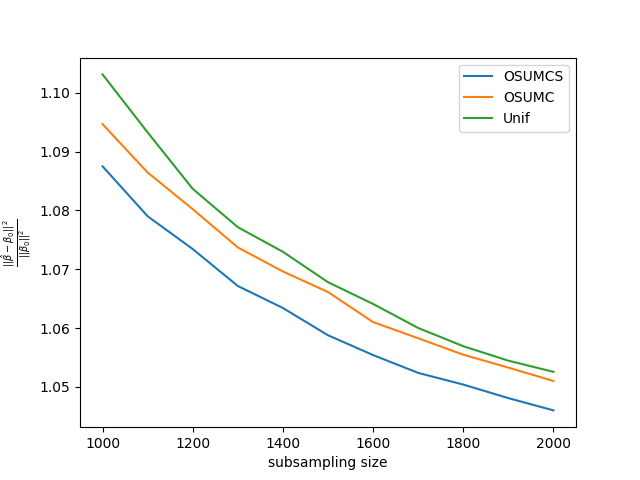}
		\caption{Estimation relative SE}  
		
	\end{subfigure}
	\begin{subfigure}[b]{0.49\linewidth}
            \centering
		\includegraphics[width=\textwidth]{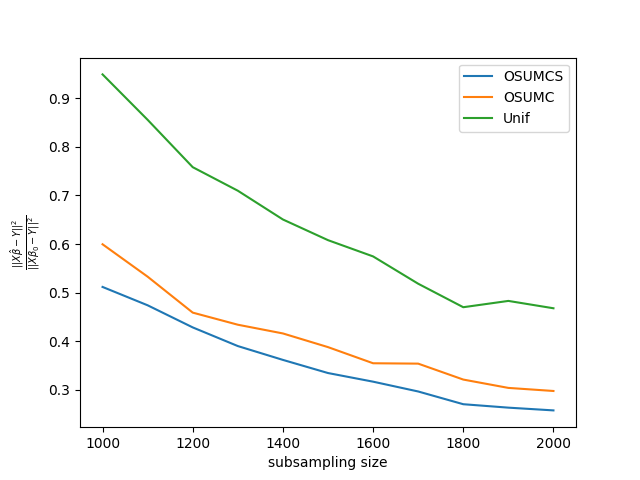}
		\caption{Prediction relative SE}
		
	\end{subfigure}
        \caption{Comparisons of the estimation and prediction RMSE results of OSUMCS, OSUMC, and Unif across varying subsample sizes, n, under four poisson regression scenarios.}
\end{figure}

For performance metrics, as in \cite{zhang2021} we calculate the relative mean squared error (RMSE) for estimation as $||\hat{\bbeta} - \bbeta_0||^2 / ||\bbeta_0||^2$, and the relative squared error for prediction as $||\bX \hat{\bbeta} - Y||^2 / ||\bX \bbeta_0 - Y||^2$, computed on the test set. This evaluation process is repeated 100 times across a range of subsample sizes, from 1000 to 2000, in 100 increments, and we calculate the mean values for each performance metric at each subsample size. The results are presented in Figure 4.
\\~\\
Figure 4 illustrates that our OSUMCS method consistently achieves lower RMSE values for both estimation and prediction across all subsample sizes from 1000 to 2000. This indicates that OSUMCS not only delivers more accurate parameter estimates but also enhances predictive accuracy, outperforming competing methods by maintaining smaller deviations from the true parameter values. These results align well with both theoretical expectations and prior simulation findings, reinforcing that OSUMCS provides improved stability and precision across a range of sample sizes and data scenarios. This consistent performance advantage underscores OSUMCS’s adaptability in practical applications.

\section{Conclusion}
Under the semi-supervised learning structure, we proposed an optimal sampling strategy OSUMCS, for the measurement constraint datasets with surrogate variable achievable. Compared to OSUMC in \cite{zhang2021}, our method maintains the unconditional framework with statistical consistency and asymptotical normality proved. We are able to obtain an  estimator with lower variance theoretically and lower test standard errors compared to OSUMC. With the common numerical set up in sampling, our scheme also shows better algorithmic robustness.
\\~\\
Some possible following questions can be raised here, for example even though $\pi_i$ doesn't have a closed form, it might be possible to have a better way to get the alternate solution. Also, the way to estimate the $E(Y^2|\bX,S)$ and $E(Y|\bX,S)$ can also be investigated, for instance when the historical data exist, whether there exists a method that can combine the information of the outcome in the pilot example with the historical data. We would leave these as potential future work.

\bibliographystyle{chicago}
\bibliography{refer}
\newpage
\section*{Appendix}

\subsection*{Notation}
Let $A \in \mathbb{R}^{p \times p}$ be a matrix, and $\|A\|$ denotes its Frobenius norm. $A$ is positive definite if and only if $A > 0$. For two positive definite matrices $B$ and $C$, we have $B > C$ if and only if $B - C$ is positive definite.

\subsection*{Proof of Theorem 1 (Consistency)}
It is proved in \cite{zhang2021} that assume the following conditions	
	\begin{enumerate}[(i)]
		\item Either $(ia)$ $b_1^{''}(\cdot)$ is almost surely bounded or $(ib)$ $\bX$ is almost surely bounded. 
		 $\sup_{\beta\in \mathcal{B}} |X^T\beta|< \infty$. 
%{\color{red} The meaning of (ib) is unclear to me since $X$ is random.  Perhaps the condition should be $P(	\sup_{\beta\in \mathcal{B}} |X^T\beta|< \infty) =1$ or there exists $K<\infty$ such that $P(	\sup_{\beta\in \mathcal{B}} |X^T\beta|< K) =1$.}
		\item $E\bX\bX^T$ is finite and $g_1(\bbeta):=E\left[\{b_1'(\bX^T\bbeta)-Y\}\bX\right]$ is finite for any $\bbeta\in \mathcal{B}$. 
		\item $\sum_{i=1}^N E\left[ \frac{\{b_1'(\bX_i^T\bbeta)-Y_i\}^2}{\pi_i} x_{ij}^2\right]=o(N^2n)$ for $1\le j\le p$ and $\bbeta \in \mathcal{B}$. 
		\item $\inf_{\bbeta:||\bbeta-\bbeta_0||\ge \epsilon}||g_1(\bbeta)||>0$ for any $\epsilon>0$. 
	\end{enumerate}
 Then $\hat{\bbeta}_n\mathop{\to}\limits^{p} \bbeta_0$.

\noindent Similarly, for any \(\bgamma \in \mathcal{A}\), where $\mathcal{A}$ is compact, we can have that assuming the following conditions	
	\begin{enumerate}[(i)]
		\item Either $(ia)$ $b_2^{''}(\cdot)$ is almost surely bounded or $(ib)$ $\bX$ is almost surely bounded.  
		\item $EXX^T$ is finite and $g_2(\bgamma):=E\left[\{b_2'(\bX^T\bgamma)-S\}X\right]$ is finite for any $\bgamma\in \mathcal{A}$. 
		\item $\sum_{i=1}^N E\left[ \frac{\{b_2'(\bX_i^T\bgamma)-Y_i\}^2}{\pi_i} x_{ij}^2\right]=o(N^2n)$ for $1\le j\le p$ and $\bgamma \in \mathcal{A}$.
		\item $\inf_{\bgamma:||\bgamma-\bgamma_0||\ge \epsilon}||g_2(\bgamma)||>0$ for any $\epsilon>0$. 
	\end{enumerate}
 Then $\hat{\bgamma}_n\mathop{\to}\limits^{p} \bgamma_0$.
Also we know that the estimator from a regular score function is consistent, which is $\hat{\bgamma}_N\mathop{\to}\limits^{p} \bgamma_0$. Here the random variables are i.i.d generated, and we have the condition  (\ref{inver}) and (\ref{consis}) in Theorem 1. Therefore, the empirical estimators $\hat\bSigma_{12}$ and $\hat\bSigma_{22}^{-1}$ are consistent with the same convergence rate $O(n^{\frac{1}{2}})$, thus the product  $\hat\bSigma_{12}\hat\bSigma_{22}^{-1}$ is also consistent we have $\hat\bSigma_{12}\hat\bSigma_{22}^{-1}(\hat\bgamma_n-\hat\bgamma_N)\mathop{\to}\limits^{p} \bSigma_{12}\bSigma_{22}^{-1}(\bgamma_0-\bgamma_0) = \mathbf{0}$. As a result, $\hat\bbeta_A=\hat\bbeta_n-\hat\bSigma_{12}\hat\bSigma_{22}^{-1}(\hat\bgamma_n-\hat\bgamma_N)\mathop{\to}\limits^{p}\bbeta_0$.

\subsection*{Proof of Theorem 2 (Asymptotic Normality)}
In \cite{zhang2021} it is shown that 
assume the following conditions,
	\begin{enumerate}[(i)]
		\item $I(\bbeta_0)=-E\left\{b_1^{''}(X^T\bbeta_0)\bX\bX^T\right\}$ is finite and non-singular.\label{Con2_1}
		\item $\sum_{i=1}^{N}E\left\{\frac{b_1^{''}(\bX_i^T\bbeta_0)^2}{\pi_i}(x_{ik}x_{ij})^2\right\}=o(N^2n)$, for $1\le k,j \le p$.
		\item $b_1(x)$ is three-times continuously differentiable for every $x$ within its domain.
		\item Every second-order partial derivative of $\xi_{\bbeta}(\bX)$ w.r.t $\bbeta$ is dominated by an integrable function $\ddot{\xi}(\bX)$ independent of $\bbeta$ in a neighborhood of $\bbeta_0$.
            %\item $\lim\limits_{n\to \infty} \sum\limits_{i=1}^rE[||\xi_{ni}||^4]=0$, \label{AN1}
		%\item $\lim\limits_{n\to \infty}E[||\sum\limits_{i=1}^{r}E[\xi_{ni}\xi_{ni}^T|\mathcal{F}_{n,i-1}]-B_n||^2]=0$, {\color{red} these two will be removed}\label{AN2}
	\end{enumerate}
we have $$(\hat{\bbeta}_n-\bbeta_0)\mathop{\longrightarrow}\limits^d N(0,I(\bbeta_0)^{-1}V(S^*_N(\bbeta_0))I(\bbeta_0)^{-1}),$$ where $I(\bbeta_0)^{-1}V(S^*_N(\bbeta_0))I(\bbeta_0)^{-1} = \bSigma_{11}.$	
\\~\\
\begin{comment}
%%%Similarly, when assume the following conditions,
	\begin{enumerate}[(i)]
		\item $I(\bgamma_0)=-E\left\{b_2^{''}(X^T\bgamma_0)XX^T\right\}$ is finite and non-singular.
		\item $\sum_{i=1}^{N}E\left\{\frac{b_2^{''}(X_i^T\bgamma_0)^2}{\pi_i}(x_{ik}x_{ij})^2\right\}=o(N^2n)$, for $1\le k,j \le p$.
		\item $b_2(x)$ is three-times continuously differentiable for every $x$ within its domain.
		\item Every second-order partial derivative of $\psi_{\gamma}(x)$ w.r.t $\gamma$ is dominated by an integrable function $\ddot{\psi}(x)$ independent of $\gamma$ in a neighborhood of $\bgamma_0$.\label{AL3}
            \item $\lim\limits_{n\to \infty} \sum\limits_{i=1}^rE[||\xi_{ni}||^4]=0$, \label{AN1}
		\item $\lim\limits_{n\to \infty}E[||\sum\limits_{i=1}^{r}E[\xi_{ni}\xi_{ni}^T|\mathcal{F}_{n,i-1}]-B_n||^2]=0$, {\color{red} these two will be removed}\label{AN2}
	\end{enumerate}
we have $$(\hat{\bgamma}_N-\bgamma_0)\mathop{\longrightarrow}\limits^d N(0,I(\bgamma_0)^{-1}V(S^*_N(\bgamma_0))I(\bgamma_0)^{-1}). $$	
Under the same conditions and a similar way of proving, we can get 
$$(\hat{\bgamma}_N-\bgamma_0)\mathop{\longrightarrow}\limits^dN(0,I(\bgamma_0)^{-1}V(S_N(\bgamma_0))I(\bgamma_0)^{-1}). $$	%%%
\end{comment}

It is stated in Lemma 1 in \cite{zhang2021} that assume the first four assumptions of this theorem, $$S_N^*(\bbeta_0)=I(\bbeta_0)(\hat{\bbeta}_n-\bbeta_0)+o_p\left(\left\|\hat{\bbeta}_n-\bbeta_0\right\|\right).$$
Because of condition (\ref{Con2_1}), can rewrite this as:
$$\hat{\bbeta}_n-\bbeta_0=I^{-1}(\bbeta_0)S_N^*(\bbeta_0)+o_p\left(\left\|I^{-1}(\bbeta_0)(\hat{\bbeta}_n-\bbeta_0)\right\|\right).$$
Here $o_p\left(\left\|I^{-1}(\bbeta_0)(\hat{\bbeta}_n-\bbeta_0)\right\|\right)$ is a more precious description compare to $o_p(1)$, we from now will replace this term by $o_p(1)$ for simplicity. 
Similarly, we can get:
$$\hat{\bgamma}_n-\bgamma_0=I^{-1}(\bgamma_0)S_N^*(\bgamma_0)+o_p(1).$$

And following a similar procedure of proving, the following equation can be deducted:

$$\hat{\bgamma}_N-\bgamma_0=I^{-1}(\bgamma_0)S_N(\bgamma_0)+o_p(1).$$

Thus,
$$\hat{\bgamma}_n-\hat{\bgamma}_N=I^{-1}(\bgamma_0)(S_N(\bgamma_0)-S_N^*(\bgamma_0))+o_p(1).$$

For any fixed $a,b \in \mathbb{R}^p$, under the conditions (\ref{finite}) and (\ref{finite2}), and $\bX_i$ are i.i.d generated, after applying CLT, the expression 
$$a^T(\hat{\bbeta}_n-\bbeta_0) + b^T(\hat{\bgamma}_n-\hat{\bgamma}_N)$$ would be asymptotically normal.  Then by applying Cramer-Wold Theorem, we can get that $(\hat{\bbeta}_n-\bbeta_0)$ and $(\hat{\bgamma}_n-\hat{\bgamma}_N)$ are jointly normal asymptotically. 
\\~\\

From the results above with the condition (\ref{inver}), and $\bbeta_0$ is a fixed true value, the linear combination of $\hat{\bbeta}_n$ and $(\hat{\bgamma}_n-\hat{\bgamma}_N)$ is still normal, we have that $\hat\bbeta_A=\hat\bbeta_n-\hat\bSigma_{12}\hat\bSigma_{22}^{-1}(\hat\bgamma_n-\hat\bgamma_N)$ is also asymptotically normal.

We define: 

$$
\text{Asym Var}\left( \begin{array}{c}
\hat{\bbeta}_n - \bbeta_0 \\
\hat{\bgamma}_n - \hat{\bgamma}_N
\end{array} \right) 
= \begin{pmatrix} 
\bSigma_{11} & \bSigma_{12} \\
\bSigma_{21} & \bSigma_{22}
\end{pmatrix} 
,$$

For $\hat\bbeta_A$, we can deduct:

$$
\text{Asym E}\left[\left( \hat{\bbeta}_A - \bbeta_0 \right)\right] = \text{E}\left[\hat\bbeta_n\right]-\text{E}\left[\hat\bSigma_{12}\hat\bSigma_{22}^{-1}(\hat\bgamma_n-\hat\bgamma_N)\right] = \bbeta_0 + \mathbf{0} = \bbeta_0,
$$ 

$$
\text{cov}\left(\left( \hat{\bbeta}_A - \bbeta_0 \right),  \left( \hat{\bbeta}_A - \bbeta_0 \right)\right)
$$$$= \text{cov}\left( \left( \hat{\bbeta}_n - \bbeta_0 \right) - \bSigma_{12} \bSigma_{22}^{-1} \left( \hat{\bgamma}_n - \hat{\bgamma}_N \right),  \left( \hat{\bbeta}_n - \bbeta_0 \right) - \bSigma_{12} \bSigma_{22}^{-1}  \left( \hat{\bgamma}_n - \hat{\bgamma}_N \right)\right)
$$

$$
= \text{cov}\left( \left( \hat{\bbeta}_n - \bbeta_0 \right),  \left( \hat{\bbeta}_n - \bbeta_0 \right)\right) - 2 \cdot \text{cov}\left( \left( \hat{\bbeta}_n - \bbeta_0 \right), \bSigma_{12} \bSigma_{22}^{-1} \left( \hat{\bgamma}_n - \hat{\bgamma}_N \right)\right)
$$

$$
+ \text{cov}\left(\bSigma_{12} \bSigma_{22}^{-1} \left( \hat{\bgamma}_n - \hat{\bgamma}_N \right), \bSigma_{12} \bSigma_{22}^{-1}  \left( \hat{\bgamma}_n - \hat{\bgamma}_N \right)\right)
$$

$$
= \bSigma_{11} - 2 \cdot \bSigma_{12} \bSigma_{22}^{-1}  \bSigma_{21} + \bSigma_{12} \bSigma_{22}^{-1}  \bSigma_{21} \bSigma_{22}^{-1} \bSigma_{22} = \bSigma_{11} - \bSigma_{12} \bSigma_{22}^{-1}\bSigma_{21},
$$ notice that $\bSigma_{21}$ is the transpose of $\bSigma_{12}$.
\\~\\
We now go to get the value of $\bSigma_{12}$, and $\bSigma_{22}$.
We have gotten:
$$\hat{\bbeta}_n-\bbeta_0=I^{-1}(\bbeta_0)S_N^*(\bbeta_0)+o_p(1), \hat{\bgamma}_n-\bgamma_0=I^{-1}(\bgamma_0)S_N^*(\bgamma_0)+o_p(1), $$$$ \hat{\bgamma}_N-\bgamma_0=I^{-1}(\bgamma_0)S_N(\bgamma_0)+o_p(1),$$

 and $\hat{\bgamma}_n$, $\hat{\bgamma}_N$, and $\hat{\bbeta}_n$ are consistent, we have:
$$
\text{cov}(\hat{\bgamma}_n - \hat{\bgamma}_N, \hat{\bgamma}_n - \hat{\bgamma}_N) = \text{cov}((\hat{\bgamma}_n - \bgamma_0) - (\hat{\bgamma}_N - \bgamma_0), (\hat{\bgamma}_n - \bgamma_0) - (\hat{\bgamma}_N - \bgamma_0))
$$
$$
= \text{cov}(I^{-1}(\bgamma_0)(S_N^*(\bgamma_0) - S_N(\bgamma_0))+o_p(1), I^{-1}(\bgamma_0)(S_N^*(\bgamma_0) - S_N(\bgamma_0))+o_p(1))
$$
$$
= I^{-1}(\bgamma_0)\text{cov}((S_N^*(\bgamma_0) - S_N(\bgamma_0)), (S_N^*(\bgamma_0) - S_N(\bgamma_0)))I^{-1}(\bgamma_0)+o_p(1).
$$

And
$$
\text{cov}(\hat{\bbeta}_n - \bbeta_0, \hat{\bgamma}_n - \hat{\bgamma}_N) = \text{cov}(\hat{\bbeta}_n - \bbeta_0, (\hat{\bgamma}_n - \bgamma_0) - (\hat{\bgamma}_N - \bgamma_0)).
$$
$$
= \text{cov}(I^{-1}(\bbeta_0)S_N^*(\bbeta_0)+o_p(1), I^{-1}(\bgamma_0)(S_N^*(\bgamma_0) - S_N(\bgamma_0))+o_p(1))
$$
$$
= I^{-1}(\bbeta_0)\text{cov}(S_N^*(\bbeta_0), (S_N^*(\bgamma_0) - S_N(\bgamma_0)))I^{-1}(\bgamma_0)+o_p(1).
$$

Therefore, 
$$\bSigma_{22} = I^{-1}(\bgamma_0)\text{cov}((S_N^*(\bgamma_0) - S_N(\bgamma_0)), (S_N^*(\bgamma_0) - S_N(\bgamma_0)))I^{-1}(\bgamma_0),$$ and $$\bSigma_{12} = I^{-1}(\bbeta_0)\text{cov}(S_N^*(\bbeta_0), (S_N^*(\bgamma_0) - S_N(\bgamma_0)))I^{-1}(\bgamma_0).$$

So $$\bSigma_{22}^{-1} = I(\bgamma_0)\text{cov}((S_N^*(\bgamma_0) - S_N(\bgamma_0)), (S_N^*(\bgamma_0) - S_N(\bgamma_0)))^{-1}I(\bgamma_0),$$
and $$\bSigma_{11} - \bSigma_{12}\bSigma_{22}^{-1}\bSigma_{21} $$$$= I(\bbeta_0)^{-1}cov(S^*_N(\bbeta_0),S^*_N(\bbeta_0))I(\bbeta_0)^{-1} - I^{-1}(\bbeta_0)\text{cov}(S_N^*(\bbeta_0), (S_N^*(\bgamma_0) - S_N(\bgamma_0)))I^{-1}(\bgamma_0)I(\bgamma_0) $$$$\cdot\text{cov}((S_N^*(\bgamma_0) - S_N(\bgamma_0)), (S_N^*(\bgamma_0) - S_N(\bgamma_0)))^{-1}I(\bgamma_0)I^{-1}(\bgamma_0)\text{cov}(S_N^*(\bbeta_0), (S_N^*(\bgamma_0) - S_N(\bgamma_0)))^TI^{-1}(\bbeta_0)$$
$$= I(\bbeta_0)^{-1}cov(S^*_N(\bbeta_0),S^*_N(\bbeta_0))I(\bbeta_0)^{-1} - I^{-1}(\bbeta_0)\text{cov}(S_N^*(\bbeta_0), (S_N^*(\bgamma_0) - S_N(\bgamma_0))) $$$$\cdot\text{cov}((S_N^*(\bgamma_0) - S_N(\bgamma_0)), (S_N^*(\bgamma_0) - S_N(\bgamma_0)))^{-1}\text{cov}(S_N^*(\bbeta_0), (S_N^*(\bgamma_0) - S_N(\bgamma_0)))^TI^{-1}(\bbeta_0)$$
$$= I(\bbeta_0)^{-1}[\text{cov}(S^*_N(\bbeta_0),S^*_N(\bbeta_0))-\text{cov}(S_N^*(\bbeta_0), (S_N^*(\bgamma_0) - S_N(\bgamma_0))) $$$$\cdot\text{cov}((S_N^*(\bgamma_0) - S_N(\bgamma_0)), (S_N^*(\bgamma_0) - S_N(\bgamma_0)))^{-1}\text{cov}(S_N^*(\bbeta_0), (S_N^*(\bgamma_0) - S_N(\bgamma_0)))^T]I^{-1}(\bbeta_0).$$

\subsection*{Proof of Theorem 3 (Optimization)}

The following lemma can be used to show 
$$I(\bbeta_0)^{-1}\text{cov}(S^*_N(\bbeta_0),S^*_N(\bbeta_0))I^{-1}(\bbeta_0)\geq I(\bbeta_0)^{-1}[\text{cov}(S^*_N(\bbeta_0),S^*_N(\bbeta_0))-\text{cov}(S_N^*(\bbeta_0), (S_N^*(\bgamma_0) - S_N(\bgamma_0))) $$$$\cdot\text{cov}((S_N^*(\bgamma_0) - S_N(\bgamma_0)), (S_N^*(\bgamma_0) - S_N(\bgamma_0)))^{-1}\text{cov}(S_N^*(\bbeta_0), (S_N^*(\bgamma_0) - S_N(\bgamma_0)))^T]I^{-1}(\bbeta_0):$$

\begin{lemma}
Let B be an $n \times n$ real, symmetric, positive definite matrix, and let 
A be any non-zero matrix. Then $AB^{-1}A^T$ is positive semi-definite.
\end{lemma}

\begin{proof}

For any vector $z$, we have:

$$
z^T \cdot A \cdot B^{-1} \cdot A^T \cdot z = (A^T \cdot z)^T \cdot B^{-1} \cdot (A^T \cdot z)
$$

Since $B^{-1}$ is positive definite, this quadratic form is non-negative, implying that $A \cdot B^{-1} \cdot A^T$ is positive semi-definite.

\end{proof}

Let $A = I(\bbeta_0)^{-1}\text{cov}(S_N^*(\bbeta_0), (S_N^*(\bgamma_0) - S_N(\bgamma_0)))$ and $B = \text{cov}((S_N^*(\bgamma_0) - S_N(\bgamma_0)), (S_N^*(\bgamma_0) - S_N(\bgamma_0)))$, so $A \cdot B^{-1} \cdot A^T$ is positive semi-definite. Thus the above inequality holds.
\\~\\
Now we start with $\bX$ and $S$ be fixed. That is, $$\text{cov}(S^*_N(\bbeta_0),S^*_N(\bbeta_0)|\bX, S) = E(S^*_N(\bbeta_0)S^*_N(\bbeta_0)^T|\bX, S) -  E(S^*_N(\bbeta_0)|\bX, S)E(S^*_N(\bbeta_0)|\bX, S)^T $$

$$E(S^*_N(\bbeta_0)S^*_N(\bbeta_0)^T|\bX, S) = E_Y(E(S^*_N(\bbeta_0)S^*_N(\bbeta_0)^T|\bX, S, Y)|\bX, S),$$ where 

$$E(S^*_N(\bbeta_0)S^*_N(\bbeta_0)^T|\bX, S, Y) = E(\sum_{i=1}^N\frac{R_i}{\pi_i}(Y_i-b_1'(\bbeta_0^T\bX_i))\bX_i(\sum_{i=1}^N\frac{R_i}{\pi_i}(Y_i-b_1'(\bbeta_0^T\bX_i))\bX_i)^T|\bX, S, Y).$$
The sample is i.i.d, so 
$$= \sum_{i=1}^N\frac{1}{\pi_i^2}(Y_i-b_1'(\bbeta_0^T\bX_i))\bX_i((Y_i-b_1'(\bbeta_0^T\bX_i))\bX_i)^TE(R_iR_i)$$
$$= \sum_{i=1}^N\frac{1}{\pi_i}(Y_i-b_1'(\bbeta_0^T\bX_i))\bX_i((Y_i-b_1'(\bbeta_0^T\bX_i))\bX_i)^T.$$
Here $R_i$ is an indicator variable, so $E(R_iR_i) = E(R_i) = \pi_i.$

$$ E_Y(\sum_{i=1}^N\frac{1}{\pi_i}(Y_i-b_1'(\bbeta_0^T\bX_i))\bX_i((Y_i-b_1'(\bbeta_0^T\bX_i))\bX_i)^T|\bX, S)$$

$$= \sum_{i=1}^N\frac{1}{\pi_i}(E(Y_i^2|\bX, S)-2b_1'(\bbeta_0^T\bX_i)E(Y_i|\bX, S) +b_1'^2(\bbeta_0^T\bX_i))\bX_i\bX_i^T.$$
\\~\\
Notice that $E(S^*_N(\bbeta_0)|\bX, S)$ doesn't contain terms with $\pi_i$, so in the following trace representation, we will use C to represent this expectation. 

So $$tr(I(\bbeta_0)^{-1}\text{cov}(S^*_N(\bbeta_0),S^*_N(\bbeta_0))I^{-1}(\bbeta_0))$$
$$= \sum_{i=1}^Ntr(\frac{1}{\pi_i}(E(Y_i^2|\bX, S)-2b_1'(\bbeta_0^T\bX_i)E(Y_i|\bX, S) +b_1'^2(\bbeta_0^T\bX_i))I(\bbeta_0)^{-1}\bX_i\bX_i^TI(\bbeta_0)^{-1} + C)$$
$$= \sum_{i=1}^N\frac{1}{\pi_i}(E(Y_i^2|\bX, S)-2b_1'(\bbeta_0^T\bX_i)E(Y_i|\bX, S) +b_1'^2(\bbeta_0^T\bX_i))||I(\bbeta_0)^{-1}\bX_i||^2 + C$$
$$= \frac{1}{n}\sum_{i=1}^N\pi_i\sum_{i=1}^N\frac{1}{\pi_i}(E(Y_i^2|\bX, S)-2b_1'(\bbeta_0^T\bX_i)E(Y_i|\bX, S) +b_1'^2(\bbeta_0^T\bX_i))||I(\bbeta_0)^{-1}\bX_i||^2 + C$$
$$ \geq \frac{1}{n} 
\sum_{i=1}^N(\sqrt{E(Y_i^2|\bX, S)-2b_1'(\bbeta_0^T\bX_i)E(Y_i|\bX, S) +b_1'^2(\bbeta_0^T\bX_i)}||I(\bbeta_0)^{-1}\bX_i||)^2 + C.$$
\\~\\
This inequality comes from Cauchy-Schwarz Inequality and the equal sign holds if and only if $\pi_i \propto \sqrt{E(Y_i^2|\bX, S)-2b_1'(\bbeta_0^T\bX_i)E(Y_i|\bX, S) +b_1'^2(\bbeta_0^T\bX_i)}||I(\bbeta_0)^{-1}\bX_i||$.
\\~\\
Now let's assume $\bX$ and $S$ are random. Then we can get 
$$\text{cov}(S^*_N(\bbeta_0),S^*_N(\bbeta_0)) = E(\text{cov}(S^*_N(\bbeta_0),S^*_N(\bbeta_0)|\bX, S)) + \text{cov}(E(S^*_N(\bbeta_0)|\bX, S))),$$
so
$$tr(I(\bbeta_0)^{-1}\text{cov}(S^*_N(\bbeta_0),S^*_N(\bbeta_0))I^{-1}(\bbeta_0)) = tr(I(\bbeta_0)^{-1}E(\text{cov}(S^*_N(\bbeta_0),S^*_N(\bbeta_0)|\bX, S))I^{-1}(\bbeta_0) + C)$$
$$= E(tr(I(\bbeta_0)^{-1}\text{cov}(S^*_N(\bbeta_0),S^*_N(\bbeta_0)|\bX, S)I^{-1}(\bbeta_0))) + C.$$
\\
For any $\bX_i,$ when $\pi_i \propto \sqrt{E(Y_i^2|\bX, S)-2b_1'(\bbeta_0^T\bX_i)E(Y_i|\bX, S) +b_1'^2(\bbeta_0^T\bX_i)}||I(\bbeta_0)^{-1}\bX_i||$, the trace of this $\bX_i$ would be minimized. Therefore, this $\pi_i$ would be one solution for minimizing the trace expectation.

\newpage

%\bibliographystyle{chicago}
%\bibliographystyle{ECA_jasa}
%\bibliographystyle{ims}
%\bibliography{spglm}

\end{document}